\documentclass[12pt,preprint]{aastex}

\newcommand{\Msol}{\rm{M}_{\sun}}

\newcommand{\de}{\mbox{DeHt\,5}}
\newcommand{\masyr}{\mbox{mas\,yr$^{-1}$}}
\newcommand{\kms}{\mbox{km\,s$^{-1}$}}

\slugcomment{Accepted for publication in the Astrophysical Journal on
2010 September 13.}

\shorttitle{Faraday Rotation in the Tail of \de}
\shortauthors{Ransom et al.}

\begin{document}
      
\title{FARADAY ROTATION IN THE TAIL OF THE PLANETARY NEBULA \de}

\author{R. R. Ransom\altaffilmark{1,2}, R. Kothes\altaffilmark{2}, M.
Wolleben\altaffilmark{2} and T. L. Landecker\altaffilmark{2}}

\altaffiltext{1}{Department of Physics and Astronomy, Okanagan
College, 583 Duncan Avenue West, Penticton, B.C., V2A 8E1, Canada}

\altaffiltext{2}{National Research Council of Canada, Herzberg
Institute of Astrophysics, Dominion Radio Astrophysical Observatory,
Box 248, Penticton, BC, V2A 6J9, Canada}

\email{RRansom@okanagan.bc.ca}

\begin{abstract}
  
We present 1420~MHz polarization images of a
$5\arcdeg\times5\arcdeg$ region around the planetary nebula (PN)
\de\@.  The images reveal narrow Faraday-rotation structures on the
visible disk of \de\@, as well as two wider, tail-like, structures
``behind'' \de\@.  Though \de\@ is an old PN known to be interacting
with the interstellar medium (ISM), a tail has not previously been
identified for this object.  The innermost tail is $\sim$3~pc long
and runs away from the north-east edge of \de\@ in a direction
roughly opposite that of the sky-projected space velocity of the
white dwarf central star, WD~2218+706.  We believe this tail to be
the signature of ionized material ram-pressure stripped and
deposited downstream during a $>$74,000~yr interaction between \de\@
and the ISM.  We estimate the rotation measure ($RM$) through the
inner tail to be $-15 \pm 5$~$\rm{rad}\,\rm{m}^{-2}$, and, using a
realistic estimate for the line-of-sight component of the ISM
magnetic field around \de\@, derive an electron density in the inner
tail of $n_e = 3.6 \pm 1.8$~$\rm{cm}^{-3}$.  Assuming the material
is fully ionized, we estimate a total mass in the inner tail of
$0.68 \pm 0.33$~$\Msol$, and predict that $0.49 \pm 0.33$~$\Msol$
was added during the PN-ISM interaction.  The outermost tail
consists of a series of three roughly circular components, which
have a collective length of $\sim$11.0~pc.  This tail is less
conspicuous than the inner tail, and may be the signature of the
earlier interaction between the WD~2218+706 asymptotic giant branch
(AGB) progenitor and the ISM.  The results for the inner and outer
tails are consistent with hydrodynamic simulations, and may have
implications for the PN missing-mass problem as well as for models
which describe the impact of the deaths of intermediate-mass stars
on the ISM.

\end{abstract}


\section{INTRODUCTION\label{intro}}

Interaction with the interstellar medium (ISM) is expected for
planetary nebulae (PNe) in a later stage of evolution
\citep*[e.g.,][]{BorkowskiSS1990}.  In the early stages, the shell of
the PN (i.e., the region where the fast wind from the hot central star
collides with the slow wind from its progenitor) has high density, and
the PN expands essentially unimpeded into the surrounding ISM.  As the
nebula expands, its density decreases, and at some point the thermal
pressure in the nebular shell drops to a value equal to the ram
pressure of the ISM.  In many cases, the magnetic pressure from the
magnetized ISM should also be considered \citep[e.g.,][]{SokerD1997}.
Equal internal and external pressures mark the proposed start of the
PN-ISM interaction.  Since all stars have some peculiar motion, and
the ISM pressure is proportional to the square of the speed of the PN
system relative to the ambient ISM, the PN-ISM interaction appears
first as a bow shock upstream from the central star.  Indeed, an
asymmetric emission structure is the observational marker for PN-ISM
interaction in old PNe with large proper motions
(\citealt*{TweedyK1996}; \citealt{Xilouris+1996};
\citealt{Kerber+2000}).

Recently, two-dimensional \citep*{VillaverGM2003} and three
dimensional (\citealt{Wareing+2006agbbow};
\citealt*{WareingZO2007main}; \citealt{Wareing+2007mira}) hydrodynamic
simulations have been used to show that the interaction between an
evolved intermediate-mass star ($\sim$1--7~$\Msol$) and the ISM likely
starts during the asymptotic giant branch (AGB) stage.  The
simulations predict that for any system moving with respect to the
ambient (warm neutral or warm ionized) ISM, an upstream bow shock is
formed between the slow, dense AGB wind and the ISM.  Additionally,
the simulations predict that material is ram-pressure stripped from
the upstream interface and deposited downstream to form a comet-like
tail behind the moving system.  These predictions are confirmed for at
least one star, namely the primary (AGB) star in the Mira~AB binary
\citep[see][]{Martin+2007,Matthews+2008}.  Far-ultraviolet
observations for this high-proper-motion system show emission from
both a bow shock and a $\sim$4-pc-long tail \citep{Martin+2007}.

The simulations of \citet{Wareing+2006agbbow,WareingZO2007main}
further predict that the AGB-ISM bow shock shapes the PN expansion
during the later stages of PN evolution, and in fact describes the
PN-ISM interaction, not as an interaction between the expanding PN
shell and the ISM, but rather a collision between the PN shell and the
pre-existing AGB-ISM bow shock.  This PN-AGB-ISM interaction scenario
(also called the ``triple wind'' model) works very well to explain the
general shaping of many PN displaying characteristics of ISM
interaction \citep{WareingZO2007main}.  Moreover, for one of the more
prominent examples of ISM interaction, PN Sh\,2-188, simulations are
able to reproduce the observed structure and brightness distribution,
and predict both the time scale for the interaction and (with the
proper motion of the central star yet undetermined) the relative speed
of the system through the ISM \citep{Wareing+2006sh188}.

In addition to the predictions and apparent successes of the
triple-wind model described above, \citet{WareingZO2007main} (and
previously \citealt{VillaverGM2003}) suggest that the model may also
solve the ``missing-mass'' problem in PNe, whereby only a small
fraction of the mass ejected during the AGB phase is observed to be
present (in ionized form) during the PN phase \citep[see,
e.g.,][]{Zhang1995}.  The simulations of Villaver et al.\ and Wareing
et al.\ suggest that during the relatively long AGB phase, a large
fraction ($\gtrsim$60\%) of the expelled mass is stripped at the
AGB-ISM interface and deposited downstream.  Moreover, this process
appears to be important for systems moving near the Galactic plane at
even modest speeds with respect to the ISM.  The detections so far of
tails behind an AGB star, Mira~AB, and two PN systems, Sh\,2-188 and
Sh\,2-68 \citep[see][]{Xilouris+1996,Kerber+2002}, suggest that
intermediate-mass stars distribute mass over much larger volumes than
predicted for PN expansion alone.  Finding other tails and, more
importantly, estimating the total mass present in a tail, is thus an
important next step in quantifying the degree to which these stars
spread processed material throughout the ISM.

The surface brightnesses of the tails behind Mira~AB, Sh\,2-188, and
Sh\,2-68 are very low.  For the two PNe, the tail emission was
detected with deep $\rm{H}\alpha$ observations, while for Mira~AB, the
$\sim$4-pc-tail emission was detected with deep, narrow-band
ultraviolet observations.  Surveys in $\rm{H}\alpha$ may, if carefully
perused, reveal other tails.  However, it is unlikely that they can be
used to trace the full mass-loss histories of interacting systems,
since the material stripped during the AGB phase will be very diffuse
and at least partially deionized.  On the other hand, the ultraviolet
emission in the Mira~AB tail, though tracing a significant fraction of
the mass-loss period for that system, suggests an unusual, and perhaps
unique, emission mechanism.  Recent observations of the neutral
component in the Mira~AB tail suggest that \ion{H}{1} imaging is a
useful tool for studying the morphology and dynamics of tails behind
evolved stars, but high-resolution \ion{H}{1} images may be limited to
the densest (i.e., innermost) regions of the tails.  A better tool for
locating and studying the extended tails behind evolved
intermediate-mass stars may be radio polarimetric observations.
Polarimetric observations at frequencies $\leq$3~GHz (wavelengths
$\geq$10~cm) are very sensitive to Faraday rotation of the diffuse
Galactic synchrotron emission \citep[e.g.,][]{Uyaniker+2003}, and have
been used previously to study the interaction region at the upstream
interface between PN Sh\,2-216 and the ISM \citep{Ransom+2008}.  In
propagating through the magnetized ISM, the polarized component of the
background emission is rotated at wavelength $\lambda$\,[m] through an
angle
\begin{equation}
\Delta\theta = RM\,\lambda^2\ [\rm{rad}],
\end{equation}
where $RM$ is the rotation measure and depends on the line-of-sight
component of the magnetic field, $B_{\|}$\,$[\mu\rm{G}]$, the thermal
electron density, $n_e$\,$[\rm{cm}^{-3}]$, and the path length,
$dl$\,$[\rm{pc}]$, as
\begin{equation}
RM = 0.81\int{B_{\|}\,n_e\,dl\ [\rm{rad}\,\rm{m}^{-2}]}.
\end{equation}
Even in the relatively low electron-density environment expected in
the tail behind a PN or AGB star \citep[$n_e \approx n_H \sim
1$~$\rm{cm}^{-3}$, see][]{WareingZO2007main}, a rotation of the
background polarization angle of $\Delta\theta \sim 5\arcdeg$ can be
expected for a system moving through the ISM near the Galactic
plane.\footnote{We assume for the calculation of $\Delta\theta$ an
observing frequency of 1420~MHz ($\lambda = 21~\rm{cm}$), i.e., the
principal frequency of the Galactic Plane Surveys (e.g.,
\citealt{Landecker+2010}; \citealt{Haverkorn+2006b}), and use values
of 2~$\mu$G and 1~pc for the line-of-sight component of the Galactic
magnetic field and the path length through the tail, respectively.}
Such a rotation will result in a Faraday-rotation signature for the
tail which is readily observable, albeit one which can be corrupted by
other, perhaps more turbulent, structures along the line of sight.

In this paper, we report the discovery of a Faraday-rotation structure
consistent with a tail behind the fast-moving PN \de\@ (=DHW\,5;
PK\,111.0+11.6).  The structure is identified in the 1420~MHz
polarization images of the Canadian Galactic Plane Survey (CGPS).
\de\@ is an old PN known to be interacting with the ISM
\citep{TweedyK1996}, but no tail has previously been associated with
this object.  In \S~\ref{spacevel}, we summarize the properties of
\de\@, and compute its space velocity relative to the local ISM.  In
\S~\ref{obs}, we describe briefly the preparation of the CGPS
polarization images for the region surrounding \de\@.  In
\S~\ref{results}, we characterize the Faraday-rotation structures at
the position of \de\@ and in the tail behind \de\@, and estimate the
$RM$s through the inner and outer portions of the tail.  In
\S~\ref{discuss}, we discuss in detail the PN and tail structures, and
estimate the electron densities, and thus hydrogen-mass densities, in
the inner and outer tails.  Finally, in \S~\ref{concl}, we give a
summary of our results and present our conclusions.

\section{THE PLANETARY NEBULA \de \label{param}}

\begin{figure}
\plotone{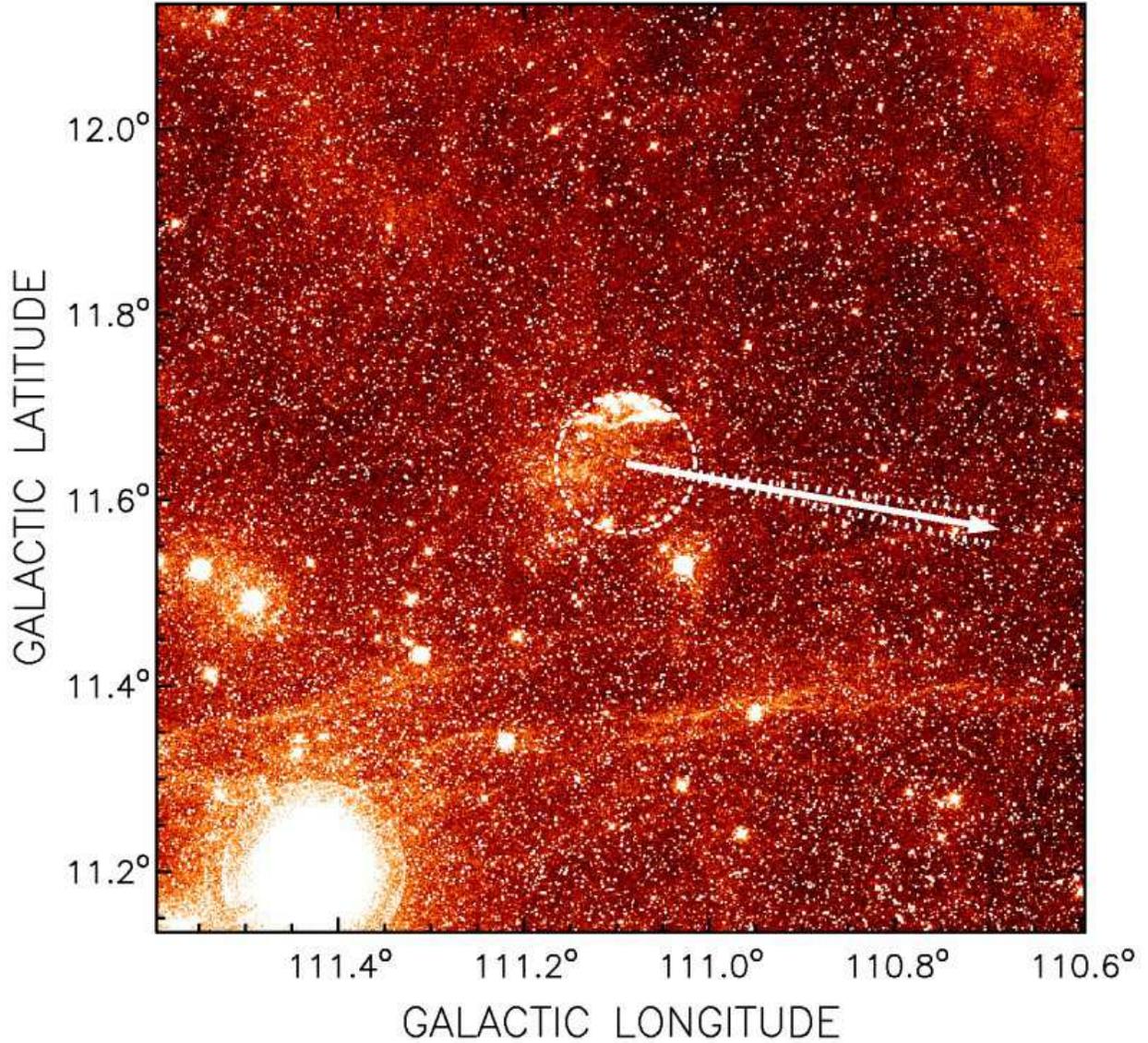}
\figcaption{R-band DSS optical image of a $1\arcdeg\times1\arcdeg$
region centered on the position of WD\,2218+706, the central star of
PN \de\@ (see Table~\ref{thestar}).  Here and hereafter, images are
presented in Galactic coordinates, with Galactic north up and Galactic
east to the left.  The intensity scale is in photon counts with
lighter shades indicating higher counts.  The range of intensities has
been adjusted to highlight extended emission.  The angular resolution
is $\sim$1\arcsec\@.  The dashed circle drawn on the image, and on
each subsequent image, shows the approximate extent of the ``visible
disk'' (see \S~\ref{param}) of \de\@.  The solid arrow and adjacent
dotted lines on the image, and on the images presented in
Figs.\,\ref{zoompolimages} and \ref{fullpolimages}, indicate the
projected space velocity and space-velocity error cone of
WD\,2218+706.  The length of the solid arrow represents the change in
the sky position of WD\,2218+706 over the next $\approx$50,000 yrs.
The filamentary structure at the northernmost edge of \de\@ is also
given the designation LBN\,538.  The bright star in the south-east
corner of the image is HR~8557 ($m_V = 5.5$).
\label{opticalimage}}
\end{figure}

\de\@ is an old PN (see Table~\ref{thestar}) whose optical morphology
clearly demonstrates interaction with the ISM.  Indeed, the
narrow-band and emission-line images of \citet{TweedyK1996} show that
the intrinsic morphology of the PN may have been largely destroyed by
the PN-ISM interaction.  For point of illustration, we show in
Figure~\ref{opticalimage} the Digitized Sky Survey (DSS) R-band
($\lambda = 657$\,nm) optical image for the $1\arcdeg\times1\arcdeg$
region centered on the \de\@ white dwarf central star, WD~2218+706.
The brightest and most distinct portion of the PN is marked by the
filamentary emission to the (Galactic) north of WD~2218+706.  The
remainder of the PN is significantly more diffuse and amorphous.  The
9$\arcmin$-diameter dashed circle in Figure~\ref{opticalimage} (and in
subsequent figures featuring the radio polarization images) represents
the approximate visible extent of \de\@ \citep[see][]{TweedyK1996},
and, for lack of a better reference point, is centered on WD~2218+706.
We emphasize, however, that the ``visible disk'' of \de\@ has neither
a distinct circular shape nor is it necessarily centered on
WD~2218+706.  In fact, given the age of \de\@, it is very likely that
WD~2218+706 is offset from the center of the PN in the direction of
motion.  We therefore advise the reader to use the disk outline as it
appears in each figure only as a rough guide to the size and position
of \de\@.

\subsection{Space Velocity of \de \label{spacevel}}

We can use the parallax, proper motion, and radial velocity of
WD~2218+706 (see Table~\ref{thestar}) to estimate the UVW motion of
\de\@ through the ISM, where the U component is positive toward the
Galactic center, V is positive in the direction of Galactic rotation,
and W is positive toward the north Galactic pole
\citep[e.g.,][]{JohnsonS1987}.  After correcting for a solar motion of
$(\rm{U}_{\sun}, \rm{V}_{\sun}, \rm{W}_{\sun}) = (10.00 \pm 0.36, 5.25
\pm 0.62, 7.17 \pm 0.38)$ $\rm{km}\,\rm{s}^{-1}$ \citep{DehnenB1998b},
we find velocity components for WD~2218+706 of $(\rm{U}, \rm{V},
\rm{W}) = (+54.18 \pm 1.28, -18.14 \pm 3.38, -15.39 \pm 0.81)$
$\rm{km}\,\rm{s}^{-1}$.  The overall space velocity of \de\@ through
the ambient\footnote{We do not take into account any motion from an
interstellar cloud that may contain \de\@.  If \de\@ is indeed part of
a cloud, then the quoted space velocity may be off by up to
$\sim$5~$\kms$ \citep[see][]{Bannister+2001}.} ISM is then $59.17 \pm
3.70$~$\kms$, with components on the sky and along the line of sight
of $45.8 \pm 2.4$~$\kms$ and $37.5 \pm 3.2$~$\kms$, respectively.  The
sky-projected space velocity of \de\@, including error cone, is
illustrated in Figure~\ref{opticalimage} (optical image) as well as
Figures~\ref{zoompolimages} and \ref{fullpolimages} (radio
polarization images).

The space velocity of \de\@ is consistent with the average value
($\sim$60~$\kms$) given by \citet{BorkowskiSS1990} for Galactic disk
PNe.  To better establish its population membership, we integrated the
Galactic orbital motion of \de\@ using the initial velocity components
given above and the Galactic gravitational potential derived by
\citet{DaupholeC1995}.  We do not plot the results, but note that the
orbit of \de\@ is confined to within 200~pc of the Galactic plane,
indicating that \de\@ is a thin disk PN.  We discuss the trajectory of
\de\@ over the last $\sim$2~Myr in \S~\ref{evidencefortail}.

\section{OBSERVATIONS AND IMAGE PREPARATION \label{obs}}

The radio polarization data presented in this paper were obtained at
1420~MHz ($\lambda = 21$\,cm) as part of the CGPS \citep{Taylor+2003}.
The particular area of interest actually falls in the high-latitude
extension of the CGPS: a region between Galactic longitudes $l =
100\arcdeg$ and $l = 117\arcdeg$ for which the nominal coverage in
Galactic latitude (i.e., $-3.6\arcdeg < b < +5.6\arcdeg$) is increased
to $-3.6\arcdeg < b < +17.5\arcdeg$ \citep[see][]{Landecker+2010}.
The principal observing instrument for the radio component of the CGPS
is the synthesis telescope \citep[ST; see][]{Landecker+2000} at the
Dominion Radio Astrophysical Observatory (DRAO).  The ST collects
continuum data in four 7.5~MHz bands centered on 1406.65, 1414.15,
1426.65 and 1434.15~MHz, respectively.\footnote{Note that the
frequency corresponding to the midpoint of the four continuum bands is
1420.4~MHz, the neutral hydrogen spin-flip frequency.  The 5.0~MHz
band about this frequency is allocated to the 256-channel spectrometer
\citep[see][]{Taylor+2003}.} The ST is sensitive at 1420~MHz to
emission from structures with angular sizes of $\sim$1\arcdeg\@ down
to the resolution limit of $\sim$1\arcmin\@.  In Stokes-Q ($Q$) and
Stokes-U ($U$), data from two single-antenna surveys of the northern
sky at $\sim$1.4~GHz, namely the DRAO-26m survey \citep{Wolleben+2006}
and the Effelsberg Medium Latitude Survey \citep{Reich+2004}, are
added to the band-averaged ST data to provide information on the
largest spatial scales \citep[see][]{Landecker+2010}.

CGPS data calibration and processing procedures are described in
detail in \citet{Taylor+2003}.  A summary of the specific procedures
used to calibrate the polarization data is given in
\citet{Ransom+2008}.  The estimated noise level in the band-averaged
images produced for the $5\arcdeg\times5\arcdeg$ region of the CGPS
presented in this paper is $\sim$0.34~$\rm{mJy}\,\rm{beam}^{-1}$
($\sim$0.045~K).

\section{RADIO POLARIZATION IMAGES OF PN \de\@ \label{results}}

In Figure~\ref{zoompolimages} we show the polarized intensity ($P =
\sqrt{Q^2+U^2-(1.2\sigma)^2}$; e.g., \citealt{SimmonsS1985}), where
the last term gives explicitly the noise bias correction) and
polarization angle ($\theta_P = \frac{1}{2}\arctan{U/Q}$) images at
1420~MHz ($\lambda = 21$\,cm) for the $1\arcdeg\times1\arcdeg$ region
about \de\@ (i.e., the same region presented optically in
Figure~\ref{opticalimage}).  The images reveal narrow polarization
structures on and immediately (to the Galactic) north-east of the
visible disk of the PN, as well as an extended polarization structure
which runs east-northeast from the edge of the visible disk to the
edge of the displayed region.  (Recall, visible disk, or ``disk'' as
it is shortened to hereafter, refers to the approximate visible extent
of \de\@; see \S~\ref{param}.)  The reduced intensity of these
structures, compared to the relatively smooth polarized emission in
all other parts of the $1\arcdeg\times1\arcdeg$ region, indicates that
their appearances are due to the effects of Faraday rotation; namely
localized beam depolarization of the background diffuse synchrotron
emission and/or background-foreground cancellation
\citep[e.g.,][]{Gray+1999}.  The coincidence of the narrow structures
with the disk of \de\@ suggests that their origin is the shell of the
PN (see \S~\ref{disk}).

To better characterize structures outside the disk of the PN, we show
in Figure~\ref{fullpolimages} the polarized intensity and polarization
angle images for a $5\arcdeg\times5\arcdeg$ region about \de\@.  The
extended Faraday-rotation structure, noted above, is the most distinct
feature in the region.  It is $\sim$0.5$\arcdeg$ long and
$\sim$0.3$\arcdeg$ wide, with a major-axis position angle (p.a.)  of
$\sim$32\arcdeg\@ (north of east).  The major axis of the structure is
nearly (counter-)aligned with the projected space velocity of \de\@
(p.a.\ = $10.5\arcdeg \pm 2.5\arcdeg$, south of west), suggesting that
the origin of this structure may be a tail (labeled ``thick tail'' in
Figure~\ref{fullpolimages}) of ionized material deposited downstream
from the PN.  We estimate the $RM$ of the thick tail in
\S~\ref{thicktail}, and provide evidence for its physical association
with \de\@ in \S~\ref{evidencefortail}.  If we track north-east from
the easternmost edge of this tail-like structure, we encounter three
additional, though more subtle, Faraday-rotation structures (labeled
``thin tail'' in Figure~\ref{fullpolimages}) along roughly the same
line.  We estimate the mean $RM$ of these structures in
\S~\ref{thintail}, and speculate on their origin in
\S~\ref{evidencefortail}.

We do not believe any of the remaining polarization structures in
Figure~\ref{fullpolimages} are associated with \de\@.  The filamentary
structure immediately south-east of \de\@ (see also
Figure~\ref{opticalimage}) is coincident with optical filaments
described by \citet{BallyR2001} and considered by them to be part of
supernova remnant (SNR) G110.3$+$11.3.

\subsection{Faraday-Rotation on the Disk of \de\@ \label{disk}}

\begin{figure}
\plotone{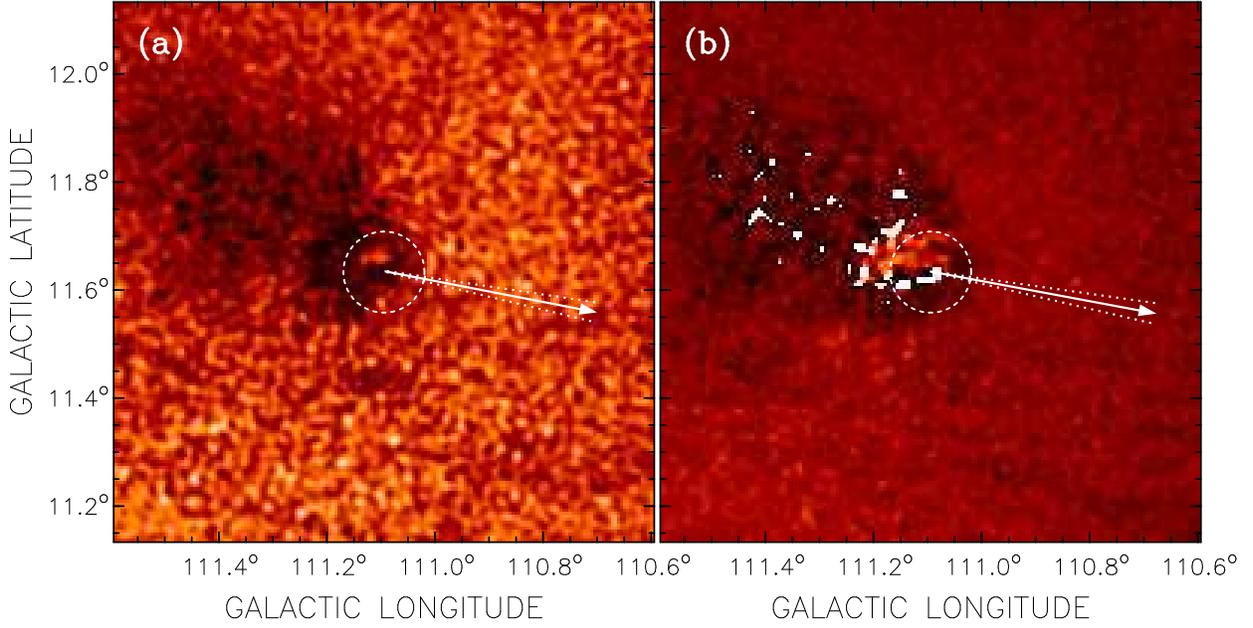}
\figcaption{Images at 1420~MHz in $(a)$ polarized intensity, $P =
\sqrt{Q^2+U^2-(1.2\sigma)^2}$, and $(b)$ polarization angle, $\theta_P
= \frac{1}{2}\arctan{U/Q}$ of the same $1\arcdeg\times1\arcdeg$ region
presented in Fig.\,\ref{opticalimage}.  The intensity scale is in
brightness temperature in $(a)$ and runs from $0$ to $0.63$\,K, with
lighter shades indicating higher temperatures.  The intensity scale in
$(b)$ extends from $-90\arcdeg$ (black) to $+90\arcdeg$ (white).  Note
that abrupt black-to-white transitions in $(b)$ do not represent large
changes in angle, since polarization angles of $-90\arcdeg$ and
$+90\arcdeg$ are equivalent.  The resolving beam at the center of each
image is $1.17\arcmin \times 1.11\arcmin$ (full-width at half-maximum;
FWHM) oriented at a position angle (east of north) of $-50\arcdeg$.
The resolving beam varies in size over each image by $\sim$3\%.
\label{zoompolimages}}
\end{figure}

\begin{figure}
\plotone{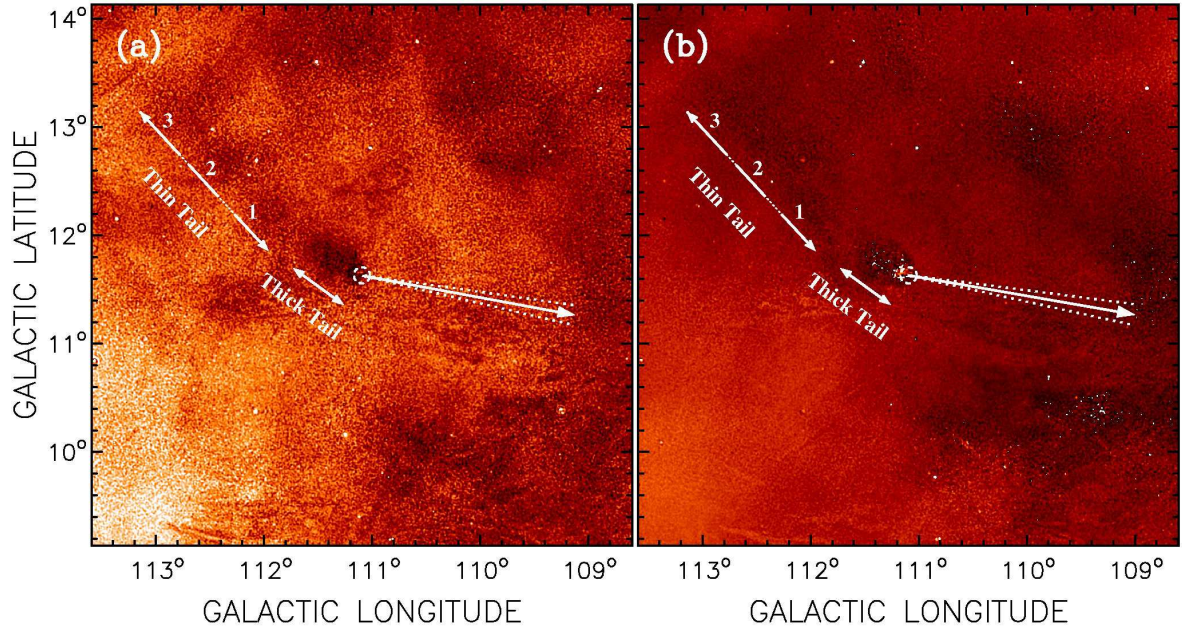}
\figcaption{Images at 1420~MHz in $(a)$ polarized intensity and $(b)$
polarization angle for a $5\arcdeg\times5\arcdeg$ region around \de\@.
The intensity scales are as described in Fig.\,\ref{zoompolimages}.
The approximate extents of the thick and thin tails described in the
text are indicated.  The three regions comprising the thin tail are
labeled 1--3, with the numbers increasing with distance from \de\@.
The length of the solid arrow represents (at this scale) the change in
the sky position of WD\,2218+706 over the next $\approx$250,000 yrs.
\label{fullpolimages}}
\end{figure}

\begin{figure}
\plotone{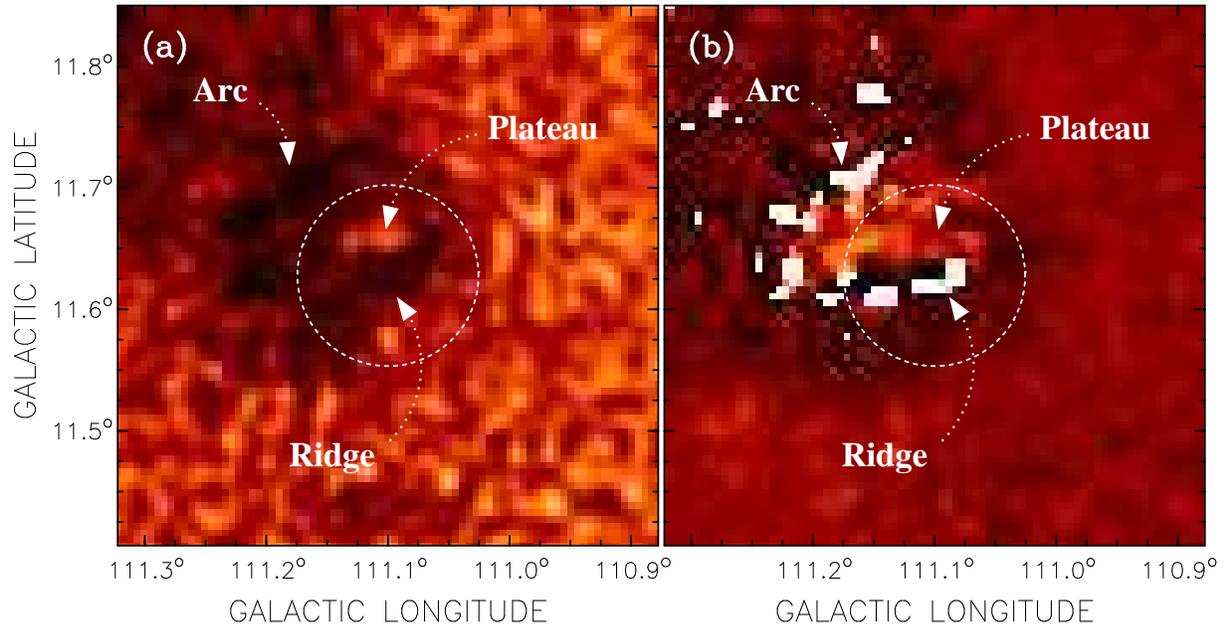}
\figcaption{$(a)$ Polarized intensity and $(b)$ polarization angle
images zoomed-in to a $0.4\arcdeg\times0.4\arcdeg$ region around
\de\@.  The intensity scales are as described for
Fig.\,\ref{zoompolimages}.  The disk features discussed in the text
are labeled.
\label{zoompa}}
\end{figure}

Figure~\ref{zoompolimages}$a$ and a close-up view in
Figure~\ref{zoompa}$a$ show a low-polarized-intensity ridge which runs
approximately east-to-west through the projected center of \de\@, and
a low-polarized-intensity arc which approximately traces the
north-east edge of \de\@.  The mean polarized intensities for the
ridge and arc are $0.103 \pm 0.044$~K and $0.081 \pm 0.039$~K,
respectively, or $\sim$0.35 and $\sim$0.27 times the off-source value
of $0.297 \pm 0.047$~K (see Table~\ref{PIandPAvalues}).  In addition,
each of these structures has small beam-sized (i.e.,
$\sim$1$\arcmin$-diameter) areas for which the polarized intensities
are virtually zero.  The low mean intensities and small-scale spatial
variations in both the ridge and arc point to beam depolarization
(rather than background/foreground cancellation) as the dominant
depolarization mechanism in these structures.  Moreover, beam
depolarization is consistent with the Faraday rotation for the ridge
and arc occurring in the shell of \de\@: sharp $RM$ gradients are
caused by fluctuations (in projected position on the sky) in either
electron density or line-of-sight magnetic field strength (see
Equation~2), each of which is expected in the turbulent shell of a PN
\cite[e.g.,][]{Ransom+2008}.

Figure~\ref{zoompolimages}$b$ and a close-up view in
Figure~\ref{zoompa}$b$ show that the polarization angles for the ridge
and arc are significantly rotated compared to the off-source value of
$-48\arcdeg \pm 6\arcdeg$ (see Table~\ref{PIandPAvalues}).  Moreover,
the polarization angles at the outside edges of the ridge and arc
(white or near-white pixels) appear to outline a single ``bent''
structure with clear boundaries on the north-east side (arc) and south
side (ridge).  At approximately the midpoint between the arc and ridge
(see Figure~\ref{zoompa}), the polarization angles plateau at
$-29\arcdeg \pm 9\arcdeg$, a value different from that seen both
off-source (i.e., north-west, west, and south of the disk) and in the
thick tail immediately north-east of the arc (see
Table~\ref{PIandPAvalues}).  Compared to the off-source value, the
mean polarization angle in the ``plateau'' is rotated $+19\arcdeg \pm
11\arcdeg$.  We infer positive, i.e., counter-clockwise, rotation,
since the alternative, namely a negative rotation of $\sim
-161\arcdeg$, would likely produce a clear (black-to-white) boundary
in polarization angle at the north and north-west edges of the disk.
No such boundary is seen.  Also, the polarization angles in our four
ST bands (see \S~\ref{obs}) do not reflect a $RM$ $\sim$9 times larger
than that derived below for the (smaller) positive rotation.
(Rotations corresponding to additional positive or negative
180$\arcdeg$ ``wraps'' in polarization angle are excluded by the
data.)  The mean polarized intensity over the plateau is $0.252 \pm
0.045$~K, a very modest reduction from the off-source value, which
suggests for this small region that background/foreground cancellation
is more important than beam depolarization.  If we assume that the
Faraday rotation takes place in the shell of \de\@, i.e., at a
distance of $345 \pm 20$~pc, and use the values for the foreground
polarized intensity and polarization angle given in \S~\ref{thicktail}
for the line of sight toward the thick tail, we estimate by comparing
off-source and on-source polarization angles a mean $RM$ over the
plateau region of $+18 \pm 12$~$\rm{rad}\,\rm{m}^{-2}$.  We comment on
the viability of using on-source/off-source polarization angles to
estimate $RM$s in \S~\ref{thicktail}.

The most prominent polarization structures on (or near) the disk of
\de\@ (i.e., the ridge and arc) do not coincide spatially with the
bright filamentary structure seen in optical emission
(Figure~\ref{opticalimage}) at the north edge of \de\@.  Nevertheless,
we believe that the approximate confinement of the ridge and plateau
region to the disk of \de\@ and close correspondence between the arc
and the north-east edge of the disk provide strong evidence that each
structure (or single common structure) is physically associated with
the shell of the PN.  We elaborate on the physical association in more
detail in \S~\ref{evidenceforshell}.

\subsection{Faraday-Rotation in the Thick Tail \label{thicktail}}

The $\sim$0.5$\arcdeg$ $\times$ $\sim$0.3$\arcdeg$ thick tail (see
Figures~\ref{zoompolimages} and \ref{fullpolimages}) stands out
noticeably in both polarized intensity and polarization angle relative
to the (off-source) region near \de\@.  The $0.162 \pm 0.055$~K mean
polarized intensity in the thick tail is $\sim$0.55 times the
off-source value of $0.297 \pm 0.047$~K, and the $-72\arcdeg \pm
8\arcdeg$ mean polarization angle is rotated $\sim$$-24\arcdeg$
compared to the off-source value of $-48\arcdeg \pm 6\arcdeg$ (see
Table~\ref{PIandPAvalues}).  A negative, i.e., clockwise, rotation
makes sense, given the absence of sharp polarized-intensity and
polarization-angle boundaries at the edge of the thick tail, and is
supported by our Faraday-rotation models (see below).  While the
reduction in mean intensity is more modest for the thick tail than for
the narrow features on the disk of \de\@, small beam-sized ``knots''
of near-zero polarized intensity and rapid polarization-angle
variation indicate the presence of sharp, localized $RM$ gradients in
the thick tail.  As with the ridge and arc, beam depolarization
probably defines the knots.

If we exclude the knots, the mean polarized intensity in the thick
tail increases slightly to $0.182 \pm 0.039$~K, or $\sim$0.61 times
the off-source value. The mean polarization angle is $-66\arcdeg \pm
6\arcdeg$, a rotation of $-18\arcdeg \pm 8\arcdeg$ from the off-source
value.  The $\sim$39\% reduction in the polarized intensity compared
to that off-source, and relatively small standard deviations about the
mean values in polarized intensity and polarization angle (compared
the ridge and arc), suggest that, outside of the knots,
background/foreground cancellation is of comparable importance to beam
depolarization.  We constructed Faraday-rotation models using the
approach of \citet{WollebenR2004}.  We treated the thick tail
(excluding the knots) as a Faraday ``screen'' that rotates the
polarization angle of the background emission.  When the rotated
background is vector-added to the polarized foreground, the net
polarized intensity drops (compared to regions outside of the screen).
For each model, we varied the $RM$ and degree of beam depolarization
for the screen as well as the foreground polarized intensity and
polarization angle.  Our set of acceptable models (judged against the
scatter in the on-source and off-source data) gives a $RM$ over the
thick tail of $-15 \pm 5$~$\rm{rad}\,\rm{m}^{-2}$.  The models also
give a foreground polarized intensity of $0.160 \pm 0.050$~K and
foreground polarization angle of $-45\arcdeg \pm 10\arcdeg$, and
indicate that beam depolarization accounts for $40\% \pm 10\%$ of the
intensity reduction in the thick tail.  The foreground polarized
intensity is consistent with the 0.12--0.23~K value suggested by
\citet{Roger+1999} for the $345 \pm 20$~pc line of sight toward
\de\@.\footnote{\citet{Roger+1999} give emissivity measurements at
22~MHz for lines of sight near \de\@.  \citet{ReichR1988} show that
the brightness temperature spectral index between 408 and 1420~MHz for
the region around \de\@ is $\beta = -2.7$ ($T_B \sim \nu^{\beta}$).
If we assume that this value applies also to the frequency range
22--408~MHz, then (assuming the emission is $\approx$70\% polarized)
the Roger et al.\ measurements give a foreground polarized intensity
of $0.125 \pm 0.010$~K.  However, low-angular-resolution maps between
38 and 408~MHz suggest a slightly flatter index \citep[$\beta =
-2.5$;][]{Lawson+1987} at lower frequencies.  Applying $\beta = -2.5$
between 22 and 408~MHz and $\beta = -2.7$ between 408 and 1420~MHz,
the Roger et al.\ measurements give a foreground polarized intensity
of $0.215 \pm 0.015$~K.}  Note that the polarization angle is very
similar to the nominal off-source value.

Estimating $RM$s by comparing on-source and off-source polarization
angles is subject to rather large uncertainties, particularly when the
foreground toward the source of interest contributes a significant
fraction of the polarized emission ($\gtrsim$50\%).  While we believe
that the value quoted above for the thick tail is accurate within the
uncertainties, a direct estimate of the $RM$ using (widely spaced)
multi-frequency polarimetric data would be superior.  Future
polarimetric observations at subarcminute resolution and at multiple
frequencies in the 1--3~GHz range could provide a detailed $RM$-map of
the region around \de\@.

\subsection{Faraday-Rotation in the Thin Tail \label{thintail}}

The less conspicuous thin tail (see Figure~\ref{fullpolimages}) is
comprised of three Faraday-rotation structures, each roughly
$\sim$0.3$\arcdeg$ in diameter, and has a total length of
$\sim$1.8$\arcdeg$.  The mean polarized intensity across all three
structures is $0.260 \pm 0.048$, or $\sim$0.84 times the off-source
value for the region around the thin tail of $0.309 \pm 0.047$~K (see
Table~\ref{PIandPAvalues}).  The mean polarization angle across all
three structures is $-55\arcdeg \pm 6\arcdeg$, a rotation of
$-7\arcdeg \pm 8\arcdeg$ from the off-source value of $-48\arcdeg \pm
6\arcdeg$ (see Table~\ref{PIandPAvalues}).\footnote{Though roughly
consistent with zero, the $-7\arcdeg \pm 8\arcdeg$
on-source/off-source rotation nevertheless yields thin-tail structures
which stand out in contrast relative to the smoother off-source
regions.  The true standard error for the rotation is likely smaller
than $8\arcdeg$ (computed as the root-sum-square of the standard
deviations of the polarization angles for the thin tail and off-source
region), but is difficult to estimate in the absence of a
Faraday-rotation model.}  We again infer a negative rotation, since
there are no sharp boundaries at the edges of these structures.  If we
assume that the Faraday rotation takes place at a distance of $345 \pm
20$~pc, and use the values for the foreground polarized intensity and
polarization angle given in \S~\ref{thicktail} for the line of sight
toward the thick tail, we estimate by comparing off-source and
on-source polarization angles a mean $RM$ over the thin tail of $-7
\pm 10$~$\rm{rad}\,\rm{m}^{-2}$.  We tried to check the consistency of
the values for the foreground polarized intensity and polarization
angle, but were not able, due to the relatively weak depolarization
signatures in the thin-tail structures, to solve simultaneously (using
the approach of Wolleben \& Reich) for the $RM$ and the foreground
parameters.

\section{DISCUSSION \label{discuss}}

\subsection{Evidence in Faraday Rotation of the Shell of \de\@ \label{evidenceforshell}}

Given the west-southwest projected space velocity of the \de\@ central
star, WD 2218+706, we might expect to see evidence in the optical
image (Figure~\ref{opticalimage}) and radio polarization images
(Figure~\ref{zoompolimages}) of a bow shock at the south-west edge of
the disk of the PN.  A close connection between enhanced optical
emission and Faraday rotation is observed at the leading edge of at
least one other old PN \citep[Sh~2-216;][]{Ransom+2008}.  Since,
however, both enhanced optical emission and Faraday rotation rely on
column density of ionized material, and a large component of the space
velocity of \de\@ lies along the line of sight (see
\S~\ref{spacevel}), a leading-edge enhancement may appear in
projection more like a ring for \de\@.  Moreover, recent hydrodynamic
simulations of the interaction with the ISM of moderately fast-moving
PNe (50--100~$\rm{km}\,\rm{s}^{-1}$) show that ``hot spots'' in the
nebular shell migrate over time downstream from the leading-edge of
the interaction, and sit in the more advanced stages of the PN-ISM
interaction at approximately the halfway point between the upstream
and downstream sides of the shell \citep[see][]{WareingZO2007main}.
If this is the case for the $\gtrsim$90,000~yr-old \de\@ (see
Table~\ref{thestar}), then the column density of ionized material in
the projected ring would be higher than for any other location in the
PN.

\begin{figure}
\plotone{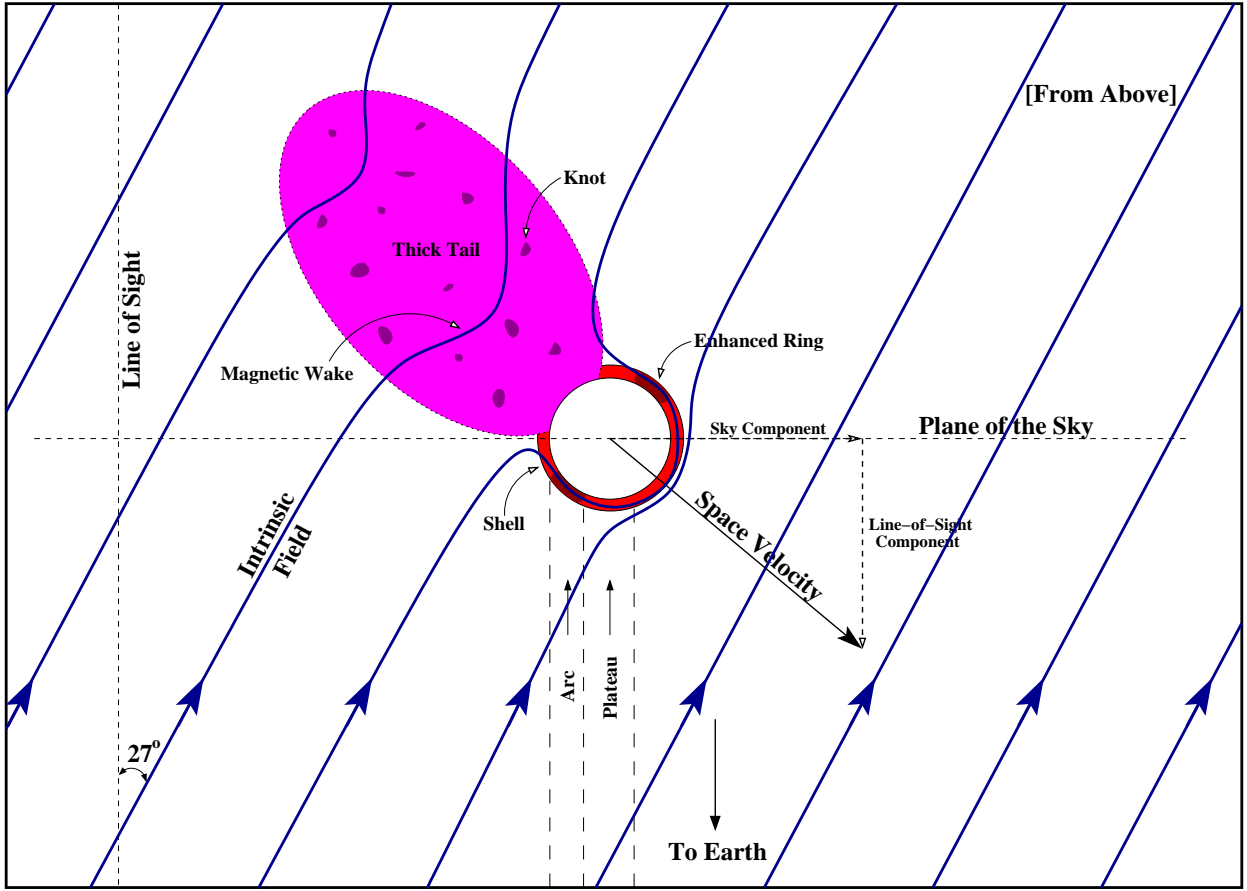}
\figcaption{Qualitative model showing the interaction between \de\@
and the magnetized ISM.  The perspective is that of an observer
sitting well above the Galactic plane and looking down on the center
of \de\@ (with Galactic east to the left). The space velocity of
WD\,2218+706, projected onto a plane parallel to the Galactic plane,
is indicated, with line-of-sight and sky components.  The intrinsic
ISM magnetic field is inclined $\approx$27$\arcdeg$ relative to the
line of sight.  The field lines immediately surrounding \de\@ are
compressed and deflected around the leading edge of the shell of the
PN.  The deflection leads to a field which has a rapidly varying
line-of-sight component on the east side of the disk (Arc) and a more
uniform orientation, with small out-of-the-sky component, near the
center of the disk (Plateau).  Note that the arc is coincident with
the front-side of the Enhanced Ring, which is oriented roughly
perpendicular to the space velocity (see \S~\ref{evidenceforshell}).
The $\sim$3-pc-long Thick Tail, representing ionized material stripped
at the interface between \de\@ and the ISM during the PN phase,
deflects the intrinsic field in the direction of motion (Magnetic
Wake), but not to the same extent as the dense PN shell.  We have
drawn the thick tail on this plane such that its major axis is aligned
with the space velocity of WD\,2218+706.  The small misalignment that
we see in the radio polarization images may lie completely in the
plane of the sky.
\label{model}}
\end{figure}

There is no indication of a ring in Figure~\ref{opticalimage}, or in
any other narrow-band or spectral-line optical image of \de\@
\citep[see][]{TweedyK1996}.  We return to this point at the end of
this subsection.  There is, however, evidence for a ring-like
structure in the radio polarization images.  The bent structure on the
disk of \de\@ (see, in particular, Figure~\ref{zoompa}$b$) may be the
outline of a ring which forms a clear boundary with the surrounding
regions on the north-east (arc) and south (ridge) sides.  Indeed, a
circular ring, aligned in three-dimensions perpendicular to the space
velocity of \de\@, would have a projected shape on the front side
(i.e., the side closest to Earth) similar to that formed by the arc
and ridge. (We show in Figure~\ref{model} a two-dimensional slice
through the ring described here on a plane perpendicular to the plane
of the sky.)  Furthermore, close inspection of the west and north-west
edges of the disk in both polarized intensity and polarization angle
shows that there is a jagged, though very narrow, transition between
the plateau and off-source region.  There are three reasons why the
ring boundary on the west side may be more subtle than that on the
east side: (1) The polarization angle difference between the plateau
and thick tail (immediately east of the disk) is larger than that
between the plateau and off-source region immediately west of the
disk.  A smaller mean off-source to on-source rotation results in a
lesser degree of beam depolarization, and consequently a thinner and
more subtle boundary. (2) There may be spherical (or circular in the
case of a ring) asymmetries in the density of material in the shell
\citep{WareingZO2007main,WareingZO2007vort}. (3) Spatial variations in
the strength and/or orientation of the magnetic field at the back
(west) side of the ring may be smaller in magnitude than those on the
front (east) side.  We elaborate on the magnetic field geometry in the
shell of \de\@ below.

We estimate in the plateau region between the ridge and arc a mean
$RM$ of $+19 \pm 11$~$\rm{rad}\,\rm{m}^{-2}$.  Since no estimate is
given in the literature for the electron density in the shell of
\de\@, we cannot derive (see eq.\ [2]) the magnitude of the
line-of-sight magnetic field in the plateau.  However, the sign of the
$RM$ indicates that the field in this central region is directed out
of the plane of the sky.  If the field is ISM in origin
\citep[e.g.,][]{Ransom+2008}, we can explain simultaneously the
out-of-the-sky component near the center of the disk of \de\@ and the
into-the-sky component discussed in \S~\ref{massinthicktail} for the
thick tail (see Figure~\ref{model}): The intrinsic ISM field in the
region about \de\@ (see \S~\ref{massinthicktail}) is deflected around
the shell of \de\@ by the fast-moving PN system, such that a small
out-of-the-sky component is generated near the center of the disk of
\de\@.  The rapid spatial change of the deflected field at the east
edge of the disk provides one ingredient for the beam depolarization
responsible for the arc.  The other ingredient is the ionized material
in the front portion of the enhanced ring.  A similar deflection and
enhancement at the bottom of the shell of \de\@ may be responsible for
the appearance of the ridge.

The narrow-band and spectral-line optical emission for \de\@ show no
clear leading edge or ring-like enhancements.  Indeed, the only
enhanced emission seen at optical wavelengths is the filamentary
emission at the north edge of the disk.  Some authors
\citep[e.g.,][]{TweedyK1996,Parker+2006} suggest that the diffuse
optical emission on the disk of \de\@ may be ionized ISM, and
speculate that the PN shell has long since dissipated.  In contrast,
the radio polarization results for both the disk and the thick tail
(see \S~\ref{evidencefortail}) indicate that \de\@, though certainly
towards the ends of its evolution, is still a PN.  Polarimetric
observations have the advantage over total-intensity observations of
being very sensitive to small spatial changes in electron density
and/or magnetic-field orientation, specifically those at the projected
boundary between the evolved shell and the off-source region.

\subsection{Two Distinct Tails in the Wake of \de\@? \label{evidencefortail}}

Are the Faraday-rotation structures behind \de\@ really the signatures
of a tail of ionized material deposited downstream by the interaction
between \de\@ (and its progenitor) and the ISM?  Could the structures
in fact trace the history of the interaction first between the AGB
wind and the ISM (thin tail) and second between the PN and the
previously established AGB-ISM interaction zone (thick tail)?  We look
in \S~\ref{thickevidence} at evidence favoring interpretation of the
thick tail as the product of material stripped during the more recent
PN-ISM interaction, and in \S~\ref{thinevidence} at evidence
supporting interpretation of the thin tail as the outcome of stripping
during the preceding AGB-ISM interaction.

\subsubsection{Evidence of the PN-ISM tail \label{thickevidence}}

The most compelling pieces of evidence that the $\sim$0.5$\arcdeg$
$\times$ $\sim$0.3$\arcdeg$ Faraday-rotation structure is the
signature of a PN-ISM tail is its north-east side attachment to \de\@
and the approximate alignment of its major axis with the projected
space velocity of WD 2218+706.  It is very unlikely that a
line-of-sight Faraday-rotation structure, unrelated to \de\@, would
have one end bounded by \de\@, and extend away from \de\@ in the
direction from which WD 2218+706 came.  But why then is the alignment
approximate, and not exact to within the error bars of the projected
space velocity?

There are two reasons why the position angle of the major axis of the
thick tail might differ by $21.5\arcdeg \pm 2.5\arcdeg$ from that of
the projected space velocity of WD 2218+706.  First, the measured
proper motion of WD 2218+706 may not represent that of the system.
Both the HST \citep{Benedict+2009HSTobsPN} and USNO
\citep{Harris+2007} observations of WD 2218+706 show signs of a binary
companion.  Using the example provided by
\citet{Benedict+2009HSTobsPN}, an M1\,V star would be at the limit of
detectability if separated from WD 2218+706 by 5.2~AU.  The resulting
11~yr orbital period for this system would result in a transverse
velocity as high as 14~$\kms$ (assuming an orbital eccentricity of
zero).  If the transverse orbital motion were approximately
perpendicular to the system proper motion, the position angle of the
projected (instantaneous) motion of WD 2218+706 would differ from that
of the system motion by $\sim$18$\arcdeg$.  A more massive K8\,V star
with a separation of just 3.2~AU would also be at the limit of
detectability \citep[see Table~8 in][]{Benedict+2009HSTobsPN}, and
would give a maximum position-angle difference of $\sim$21$\arcdeg$.
Second, the system may be accelerating toward a massive, extended
molecular cloud complex.  The cloud runs $\sim$10$\arcdeg$ from ($l =
100\arcdeg$, $b = 13\arcdeg$) to ($l = 117\arcdeg$, $b = 22\arcdeg$),
and contains $\sim$$6 \times 10^4$~$\Msol$ of material
\citep[see][]{Grenier+1989}.  (The elongated feature which runs
diagonally across the top-right portion of Figure~\ref{fullpolimages}
may be the Faraday-rotation signature of this extensive cloud.)
Distance estimates for the cloud range between $\sim$300~pc
\citep{Grenier+1989} and $\sim$400~pc \citep{BallyR2001}, indicating
that this mass is in the same region of the local arm as \de\@.  We
elaborate more on the gravitational effect of the molecular cloud on
WD 2218+706 (and its progenitor) in \S~\ref{thinevidence} below.

In addition to its location and alignment, the $\sim$0.5$\arcdeg$
major-axis length of the thick tail may itself provide evidence that
this region is the PN-ISM tail of \de\@.  At a distance of $345 \pm
20$~pc, the angular length of the thick tail corresponds to a
projected physical length of $\sim$3.0~pc.  If we assume, for the
moment, that the material in the thick tail is stationary with respect
to the ambient ISM, then the $45.8 \pm 2.4$~$\kms$ projected velocity
of WD 2218+706 leads to an age for the PN-ISM interaction of
$\sim$64,000~yr.  This value represents a lower limit, since stripped
material will not instantaneously decelerate to zero velocity.  The
simulations of \citet{Wareing+2007mira} for the interaction between
the Mira system and the ISM give a velocity lag for the tail material
of just 10--15~$\kms$, but point out that the viscosity in the tail
may be underestimated.  Taking the minimum value for the lag in the
\de\@ tail to be $\sim$10~$\kms$, we derive an upper limit for the
interaction age of $\sim$380,000~yr.  Since the PN-ISM interaction is
expected to start $\lesssim$10,000~yr after the onset of the PN phase
for fast-moving systems \citep[see][]{WareingZO2007main}, the
thick-tail length suggests that \de\@ is between 74,000 and 390,000~yr
old.  The kinematic age for \de\@ is 87,000~yr (see
Table~\ref{thestar}), close to the lower limit of this range.  In
quite some contrast, observational estimates of the mass, effective
temperature, and luminosity of WD 2218+706
\citep[see][]{Benedict+2009HSTobsPN} suggest a post-AGB age for the
transitioning star of $\gtrsim$300,000~yr \citep[see evolutionary
tracks in][]{SchonbernerB1996}, a value much closer to the upper
limit.  \citet{WareingZO2007main} suggest that the kinematic age
underestimates the true age for a PN interacting with the ISM, since
nebular expansion has ceased at the leading edge of the interaction.

The $\sim$0.3$\arcdeg$ width of the thick tail is also consistent with
the simulations of \citet{WareingZO2007main}, which show, for systems
moving at speeds of 50--75~$\kms$, a PN-ISM tail width $\sim$2 times
the diameter of the PN shell.  However, the ISM magnetic field should
also be taken into account in estimating the width of the ionized
tail.  In the case of an ISM field oriented roughly parallel to the
direction of motion, expansion of the tail material may be restricted.
In the opposite case, i.e., a field oriented perpendicular to the
direction of motion (see \S~\ref{massinthicktail}), expansion may be
facilitated.  A full magnetohydrodynamic (MHD) simulation of the
PN-ISM interaction could more clearly establish the role of the ISM
magnetic field in the evolution of the tail's morphology.  The small
beam-sized knots in the thick tail signify regions of enhanced
turbulence, and are consistent with simulations which show that
$\sim$0.1~pc vortices can be shed downstream from the leading edge of
the PN-ISM interaction \citep[see][]{WareingZO2007vort}.

\subsubsection{Evidence of the AGB-ISM tail \label{thinevidence}}

If we accept that the thick tail is the signature of the PN-ISM
interaction for \de\@, then the best evidence that the thin tail is
the signature of the earlier AGB-ISM interaction is its relative
position behind, and approximate alignment with, the thick tail.
Again, the approximate alignment requires some explanation.  The three
Faraday-rotation structures which make up the $\sim$1.8$\arcdeg$ thin
tail fall on a line with position angle $\sim$45\arcdeg\@ (north of
east).  This line is $\sim$13$\arcdeg$ more northerly than the major
axis of the thick tail.  The orbit we computed for WD~2218+706 (see
\S~\ref{spacevel}) gives a straight-line trajectory for the star over
the past $\sim$2~Myr.  However, the orbit takes into account only the
global Galactic potential, and not the contributions of close
encounters with large masses such as the molecular cloud complex
described above in \S~\ref{thickevidence}.  If the cloud is $<$100~pc
distant, the resulting acceleration would, over $\sim$1~Myr, increase
the westward velocity of the star system by $\sim$20~$\kms$, yielding
the curved trajectory between the thin and thick tails seen in
Figure~\ref{fullpolimages}.  Indeed, such an acceleration would
simultaneously explain the difference in position angles between the
thin and thick tails and between the thick tail and projected space
velocity of WD 2218+706.

At a distance of $345 \pm 20$~pc, the $\sim$1.8$\arcdeg$ angular
length of the thin tail corresponds to a projected physical length of
$\sim$11.0~pc.  Allowing the same range of velocities for the thin
tail material as for the thick tail material, we find lower and upper
limits for the age of the apparent AGB-ISM interaction of 0.23~Myr and
1.4~Myr.  The upper limit is roughly consistent with the lifetime in
the thermally-pulsing AGB (TP-AGB) phase for a star with metallicity
similar to the Sun and final core mass similar to the estimated mass
of WD~2218+706 \citep[see][]{Bloecker1995a}.  However, with seven or
more thermal pulses predicted for the full TP-AGB phase
\cite[e.g.,][]{VassiliadisW1993AGBevol}, we might expect a less
disjointed Faraday-rotation signature in the thin tail than actually
observed (see also the simulations of \citealt{Wareing+2007mira}).  On
the other hand, models suggest that mass loss is most intense (for
stars with initial mass $<$2.5~$\Msol$) during the later stages of the
TP-AGB phase, specifically the last 2--3 thermal-pulse cycles
\citep{VassiliadisW1993AGBevol,MarigoG2007}.  Indeed, the recent
models of \citet{MarigoG2007} show that the mass-loss rates are a
factor 2 or more higher for the last 2--3 pulses, and give a period
for the final pulse cycles of 0.1--0.2~Myr.  If the thin tail is the
signature of the later stages of the TP-AGB phase, then the
three-component Faraday-rotation structure suggests a time frame of
0.3--0.6~Myr, consistent with the low end of the range suggested by
the tail's full length.  A direct measurement of the velocity of the
gas in the thin tail would help to put observational constraints on
the models.

\subsection{Mass Ejected During the PN-ISM Interaction\label{massinthicktail}}

Based on the evidence and consistencies described in
\S~\ref{thickevidence}, we believe that the thick tail seen in the
radio polarization images is the Faraday-rotation signature of
material strewn downstream by the interaction between \de\@ and the
ISM.  Since we have an estimate of the $RM$ in the thick tail (see
\S~\ref{thicktail}), we can derive (via eq.\ [2]) the electron
density, and therefore mass, in the strewn material.  To estimate the
path length through the thick tail, we approximate the tail as a
cylinder with cross-sectional diameter equal to its projected
semi-minor axis ($\sim$0.3$\arcdeg$ or $\sim$1.8~pc).  The mean path
length through the thick tail is then $\Delta l_{\rm{av}} \approx
1.4$~pc.  Since there is no direct measurement of the ISM magnetic
field in the region around \de\@, we take for the nominal strength of
the field the average azimuthal value given by \citet{Heiles1996b}:
4.2~$\mu\rm{G}$.  In the local arm, this field points toward $l =
84\arcdeg \pm 4\arcdeg$ \citep[see][]{BrownT2001}.  At the midpoint of
the thick tail ($l = 111.3\arcdeg$), the azimuthal field points into
the sky, consistent with the sign of the $RM$, and makes a $27\arcdeg
\pm 4\arcdeg$ angle with the line of sight (see Figure~\ref{model}).
Considering the range of possible combinations of the regular and
random components of the Galactic magnetic field \citep[see,
e.g.,][]{Ransom+2008} along this line, we estimate an uncertainty in
the azimuthal field strength of 1.3~$\mu\rm{G}$.  With these values,
we derive for the line-of-sight component of the ISM field in the
region of the thick tail $B_{\|} = 3.7 \pm
1.3$~$\mu\rm{G}$.\footnote{We used the midpoints of the thick tail in
longitude and latitude to project the azimuthal field onto the line of
sight.  Note that, at a distance of $345 \pm 20$~pc, the thick tail
sits $\sim$70~pc above the Galactic plane, well within the predicted
$\sim$1.2-kpc scale height for the Galactic magnetic field
\citep{HanQ1994}.}  Since the azimuthal field lies largely
perpendicular to the direction of motion, and the material in the
thick tail is ionized (see below), the field is likely pulled forward
slightly inside the tail creating a ``magnetic wake'' (see
Figure~\ref{model}).  The result of this magnetic wake is a field that
has on the near (i.e., Earth-facing) side of the thick tail a
line-of-sight component smaller than the intrinsic field, and on the
far side a line-of-sight component larger than the intrinsic field.
The net effect for our purposes, assuming minimal compression of the
field lines within the tail, is a mean line-of-sight component
essentially the same as that quoted above for the intrinsic azimuthal
field.  Using ${RM} = -15 \pm 5$~$\rm{rad}\,\rm{m}^{-2}$, $\Delta
l_{\rm{av}} \approx 1.4$~pc, and $B_{\|} = 3.7 \pm 1.3$~$\mu\rm{G}$,
we derive for the electron density in the thick tail $n_e = 3.6 \pm
1.8$~$\rm{cm}^{-3}$.

The gas in the thick tail is likely completely ionized
\citep[see][]{WareingZO2007main}.  Under this assumption, the number
density for protons in the thick tail is the same as that for
electrons: $n_{\rm{H}} = 3.6 \pm 1.7$~$\rm{cm}^{-3}$.  Using a
cylindrical volume for the thick tail, with length $\sim$3.0~pc and
cross-sectional diameter $\sim$1.8~pc, we estimate the total number of
hydrogen ions in the thick tail to be $(8.1 \pm 4.0) \times 10^{56}$.
This value corresponds to a total mass of $0.68 \pm 0.33$~$\Msol$.  If
we assume that the ambient density in the region surrounding \de\@ is
$n_{\rm{H}} \sim 1$~$\rm{cm}^{-3}$, then the mass ejected into the
thick tail during the PN-ISM interaction stage is $0.49 \pm
0.33$~$\Msol$.

\subsection{Mass Ejected During the AGB-ISM Interaction\label{massinthintail}}

Using the estimate of the mean $RM$ given in \S~\ref{thintail}, we can
derive the total ionized mass in the three components which define the
thin tail.  If the thin tail is indeed the signature of the later
stages of the AGB-ISM interaction, then the ionized mass content
places a lower limit (see below) on the mass ejected during the TP-AGB
phase.  Since the uncertainty in the $RM$ for the thin-tail components
is quite large, we use only the nominal value (${RM} \sim
-7$~$\rm{rad}\,\rm{m}^{-2}$) for the following derivation, and caution
the reader to view the given ionized mass value as only a first
estimate.

To estimate the path length through the thin tail, we approximate each
component as a sphere with diameter equal to its projected diameter
($\sim$0.3$\arcdeg$ or $\sim$1.8~pc).  The mean path length through
the thin tail is then $\Delta l_{\rm{av}} \approx 1.2$~pc.  The
azimuthal ISM field (see \S~\ref{massinthicktail}) makes a $28\arcdeg$
angle with the line of sight at the midpoint of the thin tail ($l =
112.4\arcdeg$), yielding $B_{\|} \approx 3.6$~$\mu\rm{G}$.  Using
${RM} \sim -7$~$\rm{rad}\,\rm{m}^{-2}$, $\Delta l_{\rm{av}} \approx
1.2$~pc, and $B_{\|} \approx 3.6$~$\mu\rm{G}$, we derive for the
electron density in each of the thin-tail components $n_e \sim
2.0$~$\rm{cm}^{-3}$.

In contrast to the thick tail, the material in the thin tail is likely
only partially ionized.  (The ionization fraction is difficult to
estimate, since we have no independent knowledge of the ionization age
or density of the thin tail material.)  Thus, we can consider the
electron density to be only a lower limit on the mass density in the
thin-tail components: $n_{\rm{H}} \gtrsim 2.0$~$\rm{cm}^{-3}$.  Using
a spherical volume with diameter $\sim$1.8~pc for each component, we
estimate the total number of hydrogen ions/atoms in the thin tail to
be $\gtrsim$$5.4 \times 10^{56}$.  This value corresponds to a total
mass of $\gtrsim$0.45~$\Msol$.  The mass ejected into the thin tail
during the AGB-ISM interaction stage is $\gtrsim$0.23~$\Msol$, based
on an ambient density of $n_{\rm{H}} \sim 1$~$\rm{cm}^{-3}$.

\subsection{Possible Implications of the \de\@ Tails\label{implications}}

The stripped-mass estimates given above for the thick ($0.49 \pm
0.33$~$\Msol$) and thin ($\gtrsim$0.23~$\Msol$) tails, though subject
to some important assumptions and (especially for the thin tail) large
uncertainties, indicate that a significant amount of material,
originally in the envelope of the WD~2218+706 progenitor, has been
deposited downstream.  This result is consistent with hydrodynamic
simulations (\citealt{VillaverGM2003}; \citealt{WareingZO2007main}),
and has implications for (1) the PN missing-mass problem, and (2)
models which describe the impact of the deaths of intermediate-mass
stars on the ISM.  We discuss each of these briefly below.

Observations show that, on average, only 0.15~$\Msol$ of material is
present in ionized form in the shells of PNe \citep[see,
e.g.,][]{Zhang1995}.  For intermediate-mass stars with initial masses
in the 1--3~$\Msol$ range (i.e., an initial mass for which a final
core mass similar to that of WD~2218+706 is plausible), this nebular
mass represents $\lesssim$40\% of the mass lost during AGB and
post-AGB evolution \citep[e.g.,][]{Bloecker1995b}.  The ``missing''
$\gtrsim$60\% may be partly in neutral form within the PN
\citep[see][]{AaquistK1991}, but the majority is likely in the tails
of material ejected during AGB-ISM and PN-ISM interactions.  Future
radio polarimetric observations of the tail regions of \de\@ and other
PNe may yield mass estimates accurate enough to account for all of the
progenitor mass loss, and eliminate altogether the missing-mass
problem for PNe.

PNe are much smaller than SNRs: compare $\sim$2.5~pc for the average
diameter of an old PNe \citep[][]{TweedyK1996} to $\sim$30~pc for the
average diameter of a Galactic SNR \citep[][]{Kothes+2006}.  In a
one-to-one comparison, an SNR has $\sim$1730 times the volume of a PN.
If all of the mass lost by intermediate-mass stars were ultimately
``trapped'' within the shells of their PNe, then the influence of
intermediate-mass stars on the composition and dynamics of the ISM
would indeed be negligible compared to their high-mass counterparts.
The observations presented here for \de\@, however, as well as those
presented elsewhere for the Mira system and Sh\,2-188, show that the
volume of influence for intermediate-mass stars is much greater than a
PN shell.  The thick tail behind \de\@ has a volume $\sim$9 times that
of the PN.  Taking into account also the thin tail, the volume ratio
of tail to PN shell increases to $\sim$21.  If the tails behind \de\@
are representative of those behind other PNe, then the contrast
between the volume of the ISM influenced by the death of an
intermediate-mass star and that of a high-mass star is somewhat
reduced.  Furthermore, since intermediate-mass stars complete several
orbits of the Galactic center, and oscillate on their orbits up and
down through the Galactic plane, many regions of the Galactic ISM may
have been touched by their dynamically active tails.  A detailed
model, which takes into account the initial mass function as well as
lifetimes and trajectories of disk stars, is necessary in order to
fully assess the relative importance of the deaths of intermediate and
high-mass stars in the Galactic ISM.

\section{CONCLUSIONS \label{concl}}

Here we give a summary of our results and conclusions:

1.  We have presented 1420~MHz polarization images for the
$5\arcdeg\times5\arcdeg$ region around the PN \de\@.

2.  Narrow Faraday-rotation structures trace approximately the
north-east (arc) and south (ridge) edges of the disk of \de\@.  These
narrow features may represent an enhanced ring of material in the
shell of \de\@, which is aligned in three-dimensions perpendicular to
the space velocity of the white dwarf central star, WD~2218+706.

3.  A small region of relatively high polarized intensity (plateau)
sits at approximately the midpoint of the disk of \de\@.  We have
estimated via comparison of on-source and off-source polarization
angles a $RM$ through the plateau of $+18 \pm
12$~$\rm{rad}\,\rm{m}^{-2}$.  The positive sense for the $RM$
indicates that the magnetic field in the plateau is directed out of
the sky.

4.  A Faraday-rotation structure, $\sim$3~pc long and $\sim$1.8~pc
wide, runs away from the north-east edge of \de\@ in a direction
roughly opposite that of the sky-projected space velocity of the
system.  Based on its location and alignment, as well as consistencies
with hydrodynamic simulations (\citealt{VillaverGM2003};
\citealt{WareingZO2007main}), we conclude that the structure is the
signature of ionized material ram-pressure stripped and deposited
downstream during a $>$74,000~yr interaction between \de\@ and the
ISM.  We call this structure the ``thick tail'' to distinguish it from
the more subtle tail-like structure summarized in points 8--9 below.

5.  We have estimated, using the approach of \citet{WollebenR2004}, a
$RM$ through the thick tail of $-15 \pm 5$~$\rm{rad}\,\rm{m}^{-2}$.
The negative sense for the $RM$ indicates that the magnetic field is
directed into the sky.

6.  We have presented a qualitative model which interprets the
into-the-sky magnetic field in the region of the thick tail as the
large-scale azimuthal field of the Galaxy.  The out-of-the-sky
magnetic field near the center of the disk of \de\@ (see point 3
above) is naturally explained in this model as the deflected and
compressed azimuthal field.

7.  With the $RM$ given above (point 5), we have estimated the
electron density in the thick tail to be $n_e = 3.6 \pm
1.8$~$\rm{cm}^{-3}$.  Assuming that the ambient density in the region
surrounding \de\@ is $n_{\rm{H}} \sim 1$~$\rm{cm}^{-3}$, the mass
ejected into the thick tail during the PN-ISM interaction is $0.49 \pm
0.33$~$\Msol$.

8.  A series of three roughly circular Faraday-rotation structures
appear behind the thick tail, and have a collective length of
$\sim$11.0~pc.  These structures, which we call the ``thin tail,'' may
be the signature of the earlier interaction between the WD~2218+706
progenitor and the ISM.  The AGB-ISM interaction is predicted by
hydrodynamic simulations, and has been clearly observed for one system
\citep[Mira~AB; see][]{Martin+2007}.  If the thin tail behind \de\@ is
indeed the signature of the AGB-ISM interaction, then this is the
first time that both the PN-ISM and AGB-ISM interactions have been
observed for the same source.

9.  We have estimated via comparison of on-source and off-source
polarization angles a $RM$ through the thin tail of
$\sim$$-7$~$\rm{rad}\,\rm{m}^{-2}$.  With this value for the $RM$, we
have estimated the total ionized mass in the thick tail to be
$\sim$0.45~$\Msol$.  Assuming, again, that the ambient density is
$n_{\rm{H}} \sim 1$~$\rm{cm}^{-3}$, the mass ejected into the thin
tail during the $>$0.23~Myr AGB-ISM interaction is
$\gtrsim$0.23~$\Msol$. (We give a lower limit since the thin tail is
likely only partially ionized.)  Targeted, polarimetric observations
at multiple frequencies between 1 and 3~GHz would yield a more precise
estimate of the $RM$ in the thin-tail components, and allow for a more
rigorous derivation of the mass content.

10.  The discovery of the tails representing the PN-ISM and possibly
the AGB-ISM interactions for \de\@ (and its progenitor) has important
implications for the PN missing-mass problem and models which describe
the impact of the deaths of intermediate-mass stars on both the
composition and dynamics of the ISM.  Future radio polarimetric
observations of other old PNe and late-stage AGB stars may provide
additional insight.

\acknowledgements

ACKNOWLEDGMENTS.  We thank the anonymous referees for constructive
reviews of the paper and for comments helpful in the preparation of
the final manuscript.  RRR thanks the Dean of Science, Technology, and
Health at Okanagan College for arranging release time for research.
The Canadian Galactic Plane Survey is a Canadian project with
international partners, and has been supported by a grant from NSERC.
The Dominion Radio Astrophysical Observatory is operated as a national
facility by the National Research Council of Canada.  This research is
based in part on observations with the 100-m telescope of the MPIfR at
Effelsberg.  The Second Palomar Observatory Sky Survey (POSS-II) was
made by the California Institute of Technology with funds from the
National Science Foundation, the National Geographic Society, the
Sloan Foundation, the Samuel Oschin Foundation, and the Eastman Kodak
Corporation.


\bibliographystyle{apj}
\bibliography{deht5.bib}


\begin{deluxetable}{l c c c c}
\tabletypesize{\scriptsize}
\tablecaption{Properties of \de\@ and its Central Star WD\,2218+706\label{thestar}}
\tablewidth{0pt}
\tablehead{
  \colhead{Parameter} &
  \colhead{} &
  \colhead{Value} &
  \colhead{} &
  \colhead{Reference}
}
\startdata
\multicolumn{5}{c}{Nebula Properties} \\
\tableline
%
                                            &  &                                   &  & \\
Angular Diameter (arcmin)                   &  & 9                                 &  & 1 \\
Linear Diameter (pc)                        &  & $0.90 \pm 0.05$\tablenotemark{a}  &  & 1 4 \\
Kinematic Age (yrs)                         &  & 87000\tablenotemark{b}            &  & 2 4 \\
%
                                            &  &                                   &  & \\
\tableline
\multicolumn{5}{c}{Central Star Properties} \\
\tableline
%
                                            &  &                                   &  & \\
Equatorial Coordinates (J2000)              &  & 22 19 33.713, +70 56 03.28        &  & 3 \\
Galactic Coordinates (deg)                  &  & 111.093, +11.641                  &  & 3 \\
Trigonometric Parallax (mas)                &  & $2.90 \pm 0.15$\tablenotemark{c}  &  & 4 \\ 
Distance (pc)                               &  & $345^{+19}_{-17}$                 &  & 4 \\
Proper Motion ($\masyr$)                    &  & $21.93 \pm 0.12$\tablenotemark{d} &  & 4 \\
Radial Velocity ($\kms$)                    &  & $-40.9 \pm 1.5$                   &  & 5 \\
Spectral Classification                     &  & DA\tablenotemark{e}               &  & 6 \\
$T_{\rm{eff}}$ (K)                          &  & $76500 \pm 5800$\tablenotemark{f} &  & 2 \\
Mass ($\Msol$)                              &  & $0.57 \pm 0.02$                   &  & 4
\enddata
\tablecomments{Units of right ascension are hours, minutes, and
seconds, and units of declination are degrees, arcminutes, and
arcseconds; mas $\equiv$ milliarcseconds.}
\tablenotetext{a}{The linear diameter reflects the angular diameter
given by \citet{TweedyK1996} and the trigonometric parallax given by
\citet{Benedict+2009HSTobsPN}.}
\tablenotetext{b}{The kinematic age estimate is adjusted from
\citet{Napiwotzki1999} to the trigonometric parallax given by
\citet{Benedict+2009HSTobsPN}.}
\tablenotetext{c}{Estimate given is the weighted average of estimates
from the Hubble Space Telescope (HST) and the United States Naval
Observatory \citep[USNO; see][]{Harris+2007}.}
\tablenotetext{d}{Equatorial components for the proper motion:
$\mu_{\alpha} = -11.80 \pm 0.10$, $\mu_{\delta} = -18.49 \pm 0.08$
$\masyr$.}
\tablenotetext{e}{A DAO classification is used in some references,
since trace amounts of helium are present in the photosphere (see
\citealt{Barstow+2001}).}
\tablenotetext{f}{$T_{\rm{eff}}$ estimates in the literature range
between 57400~K (\citealt{Barstow+2001}) and 76500~K
(\citealt{Napiwotzki1999}).}
\tablerefs{
  1. \citealt{TweedyK1996};\phn
  2. \citealt{Napiwotzki1999};\phn
  3. \citealt{Kerber+2003};\phn
  4. \citealt{Benedict+2009HSTobsPN};\phn
  5. \citealt{Good+2005};\phn
  6. \citealt{NapiwotzkiS1995}.
  }
\end{deluxetable}

\begin{deluxetable}{l c c c c c}
\tabletypesize{\small}
\tablecaption{On-Source and Off-Source Polarized Intensities and
Polarization Angles \label{PIandPAvalues}}
\tablewidth{0pt}
\tablehead{
  \colhead{} &
  \colhead{P.A.} &
  \colhead{$\sigma_{\rm{P.A.}}$} &
  \colhead{P.I.} &
  \colhead{$\sigma_{\rm{P.I.}}$} &
  \colhead{$n$} \\
  \colhead{Region} &
  \colhead{($\arcdeg$)} &
  \colhead{($\arcdeg$)} &
  \colhead{K} &
  \colhead{K} &
  \colhead{} \\
  \colhead{(1)} &
  \colhead{(2)} &
  \colhead{(3)} &
  \colhead{(4)} &
  \colhead{(5)} &
  \colhead{(6)}
}
\startdata
\multicolumn{6}{c}{--------------------Disk Features--------------------} \\
Ridge                           & $-74$ & $15$    & $0.103$ & $0.044$ & \phn\phn410 \\
Arc                             & $-86$ & $12$    & $0.081$ & $0.039$ & \phn\phn325 \\
Plateau (between Ridge and Arc) & $-29$ & \phn$9$ & $0.252$ & $0.045$ & \phn\phn140 \\
\tableline
\multicolumn{6}{c}{--------------------Tails--------------------} \\
Thick                           & $-72$ & \phn$8$ & $0.162$ & $0.055$ & \phn2800 \\
Thick (excl.\ knots)            & $-66$ & \phn$6$ & $0.182$ & $0.039$ & \phn1650 \\
Thin                            & $-55$ & \phn$6$ & $0.260$ & $0.048$ & \phn2600 \\
\tableline
\multicolumn{6}{c}{--------------------Off-Source--------------------} \\
Disk/Thick                      & $-48$ & \phn$6$ & $0.297$ & $0.047$ & 21500 \\
Thin                            & $-48$ & \phn$6$ & $0.309$ & $0.047$ & 14000 \\
\enddata
\tablecomments{Col.\ (1) See text for description of ridge, arc,
plateau, thick tail, and thin tail.  Off-source (Disk/Thick) refers to
all pixels of the $1\arcdeg\times1\arcdeg$ region centered on \de\@
(see Figure~\ref{zoompolimages} or Figure~\ref{fullpolimages}) that
are outside the disk of \de\@ and not in the thick tail.  Off-source
(Thin) refers to all pixels within $0.5\arcdeg$ of the thin tail (see
Figure~\ref{fullpolimages}), but not in the thin tail.  Col.\ (2) Mean
polarization angle in the respective region.  (Note that polarization
angles are modulo $180\arcdeg$; i.e., angles of $-90\arcdeg$ and
$+90\arcdeg$ are equivalent.)  Col.\ (3) Standard deviation in
polarization angle.  Col.\ (4) Mean polarized intensity in the
respective region.  Col.\ (5) Standard deviation in polarized
intensity. Col.\ (6) Number of pixels used to estimate the mean
polarization angle and polarized intensity.}
\end{deluxetable}


\end{document}